\documentclass[10pt,letterpaper,twocolumn]{article}
\usepackage[english]{babel}
\usepackage{graphicx}

\newcommand{\alt}{\mathbin{\lower 3pt\hbox
   {$\rlap{\raise 5pt\hbox{$\char'074$}}\mathchar"7218$}}}
\newcommand{\agt}{\mathbin{\lower 3pt\hbox
   {$\rlap{\raise 5pt\hbox{$\char'076$}}\mathchar"7218$}}}

\begin{document}

\setcounter{footnote}{0}
\setcounter{equation}{0}
\setcounter{figure}{0}
\setcounter{table}{0}

\title{\large\bf Mechanism of Universal Conductance Fluctuations
 }

\author{ \small V. V. Brazhkin, I. M. Suslov \\
\small  Institute for High Pressure Physics, 108840
Troitsk, Moscow, Russia\\
\small  P.L.Kapitza Institute for Physical Problems,
119334 Moscow, Russia \\
{}\\
\parbox{150mm}{ \footnotesize \, Universal conductance
fluctuations  are usually observed in the form of aperiodic
oscillations in the magnetoresistance of thin wires as a function
 of the magnetic field $B$. If such oscillations are completely
 random at scales exceeding $\xi_B$, their Fourier analysis
 should reveal a white noise spectrum at frequencies below
 $\xi_B^{-1}$.
 Comparison  with the results for 1D systems suggests another scenario: according to it,
 such
 oscillations are due to the superposition of incommensurate
 harmonics and their spectrum should contain discrete
 frequencies. An accurate Fourier analysis of the classical
 experiment by Washburn and Webb reveals a purely discrete
 spectrum in agreement with the latter scenario.  However, this
 spectrum is close in shape to the discrete white noise spectrum
 whose properties are similar to a continuous one.  } }

\date{}
\maketitle

\textwidth 6.4 in
\textheight 8.5 in

\setcounter{footnote}{0}
\setcounter{equation}{0}
\setcounter{figure}{0}
\setcounter{table}{0}

Universal conductance fluctuations
[1,2] are usually observed in the form of aperiodic
 oscillations in the magnetoresistance of thin wires as a
 function of the magnetic field $B$  \cite{5} (Fig. 1).
 According to the theory [1,2], the
 conductance $G(B)$ at a given magnetic field $B$ undergoes
 fluctuations of the order of 
 $e^2/h$ under the variation of the impurity
 configuration; fluctuations in $G(B)$ and $G(B+\Delta B)$ are
 statistically independent if $\Delta B$ exceeds a certain
 characteristic scale $\xi_B$.  It is reasonable to expect that
 oscillations in $G(B)$ are completely random at
 scales exceeding $\xi_B$. Then, their Fourier analysis should
 reveal a white noise spectrum (i.e., frequency-independent
 plateau) at frequencies below $\xi_B^{-1}$.

Comparison with the results for 1D systems \cite{6} suggests another scenario.
A magnetic field perpendicular to a thin wire creates
 a quadratic potential along this wire \cite{7}, which effectively restricts the length of the system
 $L$; hence, the variation of the magnetic field is similar to
 the variation of $L$. The resistance $\rho$ of a one-dimensional
 system is a strongly fluctuating quantity and the form of its
 distribution function $P(\rho)$ strongly depends on first
 several moments. Indeed, the Fourier transform
 of $P(\rho)$
 specifies the characteristic function
$$
{F}(t)=\left\langle e^{i\rho t} \right\rangle
=\sum_{n=0}^{\infty} \frac{(it)^n}{n!}\left\langle \rho^n
\right\rangle \,,
\eqno(1)
$$
 which is the generating function of the moments $\left\langle
 \rho^n \right\rangle$. If all moments of the distribution are
 known, the function ${F}(t)$ can be constructed using them, and
 the function $P(\rho)$ is then determined by the inverse Fourier
 transform. If an increase in the moments $\left\langle \rho^n
 \right\rangle$ with $n$ is not too fast, the contributions of
 higher moments are suppressed by a factor of $1/n!$, whereas
 first several moments are significant. These moments are
 oscillating functions of $L$,
$$
\left\langle \rho \right\rangle=a_1(L)+b_1(L)\cos(\omega_1
L+\varphi_1),
\eqno(2)
$$
$$
\left\langle \rho^2 \right\rangle=a_2(L)+b_2(L)\cos(\omega_2
L+\varphi_2)+
$$
$$+
b_3(L)\cos(\omega_3
L+\varphi_3)\,, \quad\mbox{etc.,}
$$
where $a_k(L)$ and $b_k(L)$ are monotonic functions. The
reason is that the growth exponent for $\left\langle \rho^n
\right\rangle$ is determined by the $(2n\!+\!1)$th order
 algebraic equation (see Appendix), one of whose
 root is always real, whereas
 the other roots are complex
for energies in the allowed band.
 Consequently, there are $n$ pairs of complex conjugate
 roots, which ensure the presence of $n$ frequencies in
 oscillations of $\left\langle \rho^n \right\rangle$. The
 frequencies $\omega_k$
 are usually incommensurate,
 but their incommensurability vanishes
in the deep of the allowed band
 at weak disorder. According to this picture, oscillations in
 $G(B)$ shown in Fig. 1 are determined by the superposition of
 incommensurate harmonics and their Fourier spectrum should
 contain discrete frequencies. This picture is indirectly
 confirmed by the experimental data obtained in \cite{8} and
 cited in \cite{6}, according to which the distribution function
 $P(\rho)$ is not stationary, but demonstrates systematic
 aperiodic variations.

\begin{figure}
\centerline{\includegraphics[width=3.0 in]{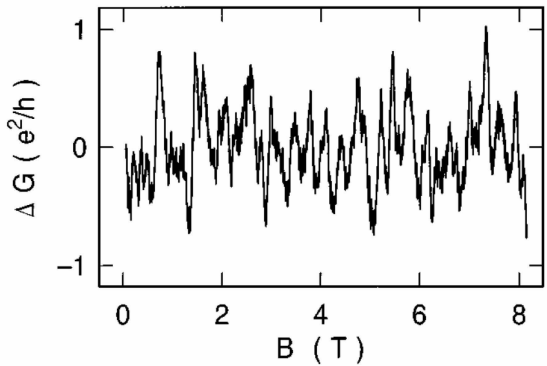}}
\caption{ Magnetic field dependence of the
conductance of the thin Au wire \cite{5}.  } \label{fig1}
\end{figure}

 It is clear from the above that the Fourier analysis of the
 function $G(B)$ makes it possible to establish which of two
 scenarios is true. However, the dependence $G(B)$ shown in Fig.
 1 cannot be used directly because a sharp cutoff of experimental
 data results in the appearance of slowly decaying oscillations
 in its Fourier transform and chaotization of the
 spectrum\,\footnote{\,Figure 14 in \cite{5} shows the Fourier
 spectrum of a thin wire in comparison with the spectrum of a
 small ring; the latter contains additional oscillations caused
 by the Aharonov--Bohm effect. However, aperiodic oscillations
were not discussed in this place
 and their spectrum, which is
 chaotic because of the sharp cutoff, was roughly approximated by
 the authors in the form of the envelope of oscillations. This is
 obvious from comparison with Figs. 12 and 13 in \cite{5}, where
 chaotic oscillations are clearly seen.}. To obtain explicit
 results, it is necessary to use an appropriate smoothing
 function.

 Let the function $f(x)$ be the superposition of discrete
 harmonics and be real. Then,
$$
f(x)=\sum_{k} A_k {\rm e}^{i\omega_k x}=
\frac{1}{2}\sum_{k} \left[ A_k {\rm e}^{i\omega_k x}
+A_k^* {\rm e}^{-i\omega_k x} \right] \,,
\eqno(3)
$$
 where the frequencies $\omega_k$ can be considered as positive
 without loss of generality. Then, the Fourier transform
of $f(x)$
 has the form
$$
F(\omega)=\pi \sum_{k} \left[ A_k \delta(\omega+\omega_k)
+A_k^* \delta(\omega-\omega_k) \right]  \,,
\eqno(4)
$$
and its modulus
$$
|F(\omega)|=\pi \sum_{k} |A_k|\left[
\delta(\omega+\omega_k) +\delta(\omega-\omega_k) \right]
\eqno(5)
$$
 depends only on the intensities of spectral lines and does not
 contain information on phase shifts in
 the corresponding harmonics.
 Since $|F(\omega)|$ is an even function,
 it is possible
  to consider
 only positive $\omega$ values and to omit the first delta
 function in Eq. (5).

\begin{figure*}
\centerline{\includegraphics[width=5.0 in]{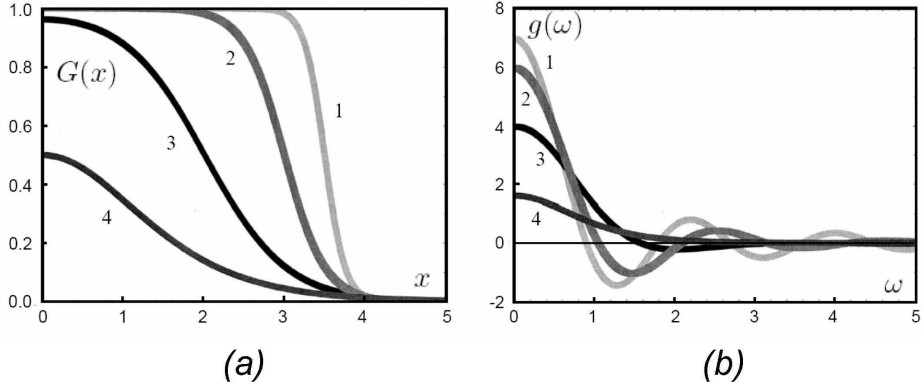}}
\caption{(a)  Function $G(x)$ given by Eq. (9) and (b)
its Fourier transform $g(\omega)$  at ({1}) $\mu=3.5$,
$T=0.125$; ({2}) $\mu=3$, $T=0.25$; ({ 3}) $\mu=2$,
$T=0.5$; and ({ 4}) $\mu=T\ln 2$, $T=0.8$.
} \label{fig2}
\end{figure*}

 Since the function $f(x)$ can be experimentally measured only in
 a certain finite $x$ range, we in practice have
$$
f(x)=\frac{1}{2}\sum_{k} \left[ A_k {\rm e}^{i\omega_k x}
+A_k^* {\rm e}^{-i\omega_k x} \right] G(x)\,,
\eqno(6)
$$
 where the function $G(x)$ is unity within the working
 range
 and zero beyond it; further, it will be smoothed. Then, instead
 of Eq. (4), we obtain
$$
F(\omega)=\frac{1}{2} \sum_{k} \left[ A_k g(\omega+\omega_k)
+A_k^* g(\omega-\omega_k) \right]   \,,
\eqno(7)
$$
 where $g(\omega)$ is the Fourier transform of
 $G(x)$, which  is real for even function $G(x)$.
   Thus, the restriction of the working
 range
 leads to the replacement of delta functions by
 spectral lines with finite widths. If discrete frequencies are
 well separated and the function $g(\omega)$ is strongly
localized
 near zero, one can neglect the overlapping of
 functions $g(\omega\pm\omega_k)$ and write at positive
 frequencies
$$
|F(\omega)|^2\approx\frac{1}{4} \sum_{k}  |A_k|^2
g^2(\omega-\omega_k)  \,.
\eqno(8)
$$
 It is preferable to use the function $|F(\omega)|^2$ (so-called
 power spectral density \cite{9}) because the integral of this
 function over all frequencies
 is equal to the  integral of $f^2(x)$ over
 all $x$ values.
 Consequently, change in  the spectrum of $f(x)$ at fixed
 rms fluctuations
 results in the redistribution of intensities between different
 frequencies at the conservation of the total spectral power.

 It is easy to see that, to obtain a clear picture in the case of
 a discrete spectrum, it is necessary to have a possibly narrower
 shape of spectral lines
 determined by $g(\omega)$,
  which can be achieved
 by the appropriate choice of the function $G(x)$. The general
 strategy is determined by the properties of integrals of rapidly
 oscillating functions \cite{10}. If
 the function $f(x)$ has discontinuity,
 its Fourier transform decreases at high frequencies as
 $1/\omega$; if the $n$th derivative is discontinuous,
 then $F(\omega)\sim \omega^{-n-1}$.  The Fourier transform of a
 smooth function $f(x)$  is calculated by shifting the contour of
 integration to a complex plane and is determined by the nearest
 singularity or saddle point, which leads to the dependence
 $F(\omega)\sim \exp(-\alpha \omega)$. If the regular function is
 obtained by means of a weak smoothing of
 a singularity, the $\alpha$ value is small and
 the exponential is manifested
 only at very high frequencies, whereas the behavior
 corresponding to the singularity holds in the remaining region.
 In our case, it is necessary to smooth the
 discontinuity  of $G(x)$.  It should be clear that weak
 smoothing is  inefficient, while strong smoothing leads to small
 values of $G(x)$ near the boundaries of the working range and to
 loss of experimental information; so, a reasonable compromise is
  required.

Let  $G(x)$ be the $x$-symmetrized Fermi
function
$$
G(x)=\frac{1}{1+{\rm e}^{(x-\mu)/T}+{\rm e}^{(-x-\mu)/T}}
=
$$
$$=
\frac{1}{1+2{\rm e}^{-\mu/T}{\rm cosh}(x/T)} \,,
\eqno(9)
$$
whose
Fourier transform is given by the integral
$$
g(\omega)=\int\limits_{-\infty}^{\infty}\frac{{\rm e}^{i\omega x}\,dx}
{b\,{\rm cosh}\beta x +c}= \frac{2\pi}{b\beta{\rm sinh} x_0}
\frac{\sin(\omega x_0/\beta)}{{\rm sinh} (\omega\pi/\beta)}\,,
$$
$$
\qquad x_0={\rm arccosh}(c/b)  \,.
\eqno(10)
$$
 If $x=B-\mu_0$ is chosen in our case, experimental data
 correspond to the interval
 $|x|\le \mu_0$ with $\mu_0=4$ (in units
 of tesla). We accepted $\mu=\mu_0-4T$, which ensures the small
 value $G(\mu_0)\approx 0.02$ at
 boundaries of the interval.
 As clear from Fig. 2,
 the behavior $g(\omega)=2\sin{\mu\omega}/\omega$
 characteristic of the sharp cutoff
 prevails
 at small $T$  values (lines {1} and {2}). It seems
 reasonable to choose $\mu=2$ and $T=0.5$
 (line {3}); in this case, 50\% of experimental data are
 effectively used, while
 the lineshape is approximately the same as in the case of
 $\mu=T\ln{2}$, where $x_0=0$,   $g(\omega)=2\pi T^2\omega/{\rm
 sinh}{\pi T\omega}$ and oscillations disappear completely (line
 {4}).

 The results of processing experimental data (Fig. 1) with the
 indicated smoothing function are shown in Fig. 3. The spectrum
 obviously consists of discrete lines, which confirms the
 second scenario given in beginning\,\footnote{\,The number and intensity of lines suggest that four 
first moments of $\rho$ are really important, the fifth moment is 
less significant, while the higher moments are practically irrelevant. }.
 However, the spectrum in the
 range $\omega\alt 2\pi/\xi_B$ (where $\xi_B$ was estimated as
 the average distance between neighboring maxima or minima in
 Fig. 1)\,\footnote{\,Under processing, Fig. 1 was strongly
 magnified and digitized by hand.
 It was revealed
 that sharp spikes in Fig. 1 are due to vertical dashes
 indicating uncertainty of the data,
 whereas the   experimental dependence is in fact smooth.} is
 similar to discrete white noise: in a rough approximation, the
 lines are equidistant and their intensities are more or less
 the same.  Since the sum over frequencies is often approximated
 by an integral, discrete white noise does not differ in many
 properties from continuous white noise.
Let, for example,
$$
F(\omega)=\pi \sum_{k} \left[ A_k \delta(\omega+\omega_k)
+A_k^* \delta(\omega-\omega_k) \right] H(\omega) \,,
\eqno(11)
$$
 where the frequencies $\omega_k$ are equidistant
 ($\omega_{k+1}-\omega_k=\Delta$), the amplitudes $A_k$ are the
 same in modulus
 ($|A_k|=A$) and have completely random phases,
 while
 $H(\omega)$ is an even function restricting the
 spectrum to the range $|\omega|\alt \Omega$.  Then, determining
 $f(x)$ by means of the inverse Fourier transform, we obtain the
 correlation function
$$
\langle f(x)f(x')\rangle=\frac{1}{2}\sum_{k} A^2 H^2(\omega_k)
{\rm e}^{i\omega_k (x-x')} \approx
$$
$$\approx
\frac{1}{2} A^2 \Delta h(x-x')    \,,
 \eqno(12)
$$
  where $h(x)$ is the Fourier transform
 of  $H^2(\omega)$.
 If the function $H(\omega)$  is smooth, $h(x)$
 decreases exponentially at a scale of $\Omega^{-1}$ in agreement
 with the diagrammatic results obtained in [1,2].

 In conclusion,
 the results obtained in this work reconcile
 two  alternative scenarios described
 at the beginning.
 On the one hand, the spectrum is discrete, confirming the
 second scenario,
 where aperiodic conductance oscillations
 are due to the superposition of incommensurate
 harmonics. On the other hand, the spectrum as a whole
 resembles
 discrete white noise, which is close in properties to
 continuous white noise.
 Universal conductance fluctuations are observed in a lot of
 works (see \cite{11,12} and references therein), and it would be
 interesting to process the corresponding experimental data in
 the spirit of the present paper.

\begin{figure}
\centerline{\includegraphics[width=3.0 in]{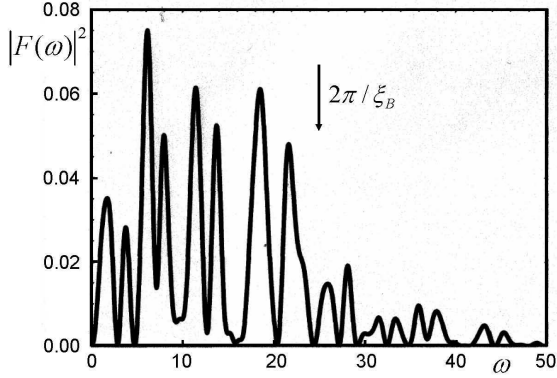}}
\caption{ Fourier analysis of the experimental data
shown in Fig. 1 with the smoothing function (9) at $\mu=2$,
$T=0.5$.
} \label{fig3}
\end{figure}

\begin{center}
{\it Appendix.} {\bf Derivation of Eq. (2)} \end{center}

Let us
consider the one-dimensional Anderson model specified by
the discrete Schr$\ddot{o}$dinger equation
$$
\Psi_{n+1}+\Psi_{n-1}+V_n \Psi_n = E \Psi_n \,,
\eqno(A.1)
$$
 where $E$ is the energy measured from the center of the band,
 $V_n$ are independent random variables with zero mean
 and  variance $W^2$, while
 the hopping integral is taken to be
 unity. Rewriting Eq. $(A.1)$ in the form
$$
\left ( \begin{array}{cc} \!\! \Psi_{n+1} \!\!
\\ \!\!\Psi_n \!\!\end{array} \right)\,
= \left ( \begin{array}{cc}\! E-V_n & -1 \! \\
 \! 1 & 0 \! \end{array} \right)\,
\left ( \begin{array}{cc}\!\! \Psi_n\!\!
\\ \!\! \Psi_{n-1}\!\!
\end{array}
\right) \,,
\eqno(A.2)
$$
and performing $n$ iterations, one can easily obtain
$$
\left ( \begin{array}{cc} \Psi_{n+1}
\\ \Psi_n \end{array} \right)\,
= \left ( \begin{array}{cc} \tau_{11} & \tau_{12}\\
\tau_{21} & \tau_{22} \end{array} \right)\,
\left ( \begin{array}{cc} \Psi_1 \\ \Psi_0 \end{array}
\right) \,.
\eqno(A.3)
$$
 Here, the matrix ${\tau}=||\tau_{ij}||$ is the product of $n$
 matrices of the form $(A.2)$ and satisfies an obvious recurrence
 relation, which can be represented in terms of matrix elements,
$$
y_{n+1}=(E-V_n) y_n +z_n\,,\qquad
z_{n+1}=- y_n \,,
\eqno(A.4)
$$
 where $y_n=\tau^{(n-1)}_{12}$, $z_n=\tau^{(n-1)}_{11}$ or
 $y_n=\tau^{(n-1)}_{22}$, $z_n=\tau^{(n-1)}_{21}$. It is
 substantial that $y^{(n-1)}$ and $z^{(n-1)}$ do not contain the
 quantity $V_{n}$ and can be averaged independently of it.
Setting $w^{(1)}_n=\langle y_n^2 \rangle$,
$w^{(2)}_n=\langle y_n z_n \rangle$,
$w^{(3)}_n=\langle z_n^2 \rangle$ for the second moments,
one can easily come to the equation
%
%
$$
\left ( \begin{array}{ccc} w^{(1)}_{n+1} \\
w^{(2)}_{n+1} \\ w^{(3)}_{n+1}
\end{array} \right)\,
= \left ( \begin{array}{ccc} E^2+W^2 &  2E &\,\,\, 1\\
-E & -1 & \,\,\, 0 \\
1 & 0 &\,\,\, 0   \end{array} \right)\,
\left ( \begin{array}{ccc}
w^{(1)}_{n} \\ w^{(2)}_{n} \\
w^{(3)}_{n}
 \end{array} \right) \,,
\eqno(A.5)
$$
Its solution is exponential,  $w^{(i)}_{n}\sim\lambda^n $,
where $\lambda$ is an eigenvalue of the matrix.
Setting $\lambda=1+x$, it is easy to
obtain an equation for $x$, which has the following form in the
limit of the continuous Schr$\ddot{o}$dinger equation:

$$
x\left(x^2 +4{\cal E}\right)= 2W^2
\eqno(A.6)
$$
 where $\cal E$ is the energy measured from the band edge. The
 equation for the growth exponent of the fourth moments can be
 obtained similarly:
$$
x\left(x^2+{4\cal E} \right)\left(x^2 +16{\cal E}\right)=
42W^2 x^2+96 W^2 {\cal E}  \,.
\eqno(A.7)
$$
 The structure of equations for arbitrary $2n$th moments can be
 established using argumentation presented in Section 4 in
 \cite{6}, where a slightly
 different formalism was used. Deep in
 the allowed and forbidden bands, only diagonal elements can be
 retained in matrices (43) and (47)  in
 \cite{6} and their analogs
 for higher moments. As a result, we arrive at the equation
$$
\prod_{k=0}^{2n}  \left[x+(n\!-\!k)\delta-B_n^k
\epsilon^2\right] = O(\epsilon^4)\,,
\eqno(A.8)
$$
where $\epsilon^2=W^2/4{\cal E}$, $\delta^2=-{\cal E}$,
$B_n^k=n(2n\!-\!1)+3k(k\!-\!2n)$.
A similar equation near the band edge
$$
x^{2n+1}=\sum\limits_{k=0}^{k_{max}} C_k W^{2k} x^{2n+1-3k}\,,
\quad k_{max}=\left[\frac{2n+1}{3} \right]
\eqno(A.9)
$$
 follows from observation that all terms of the equation
 have the same order of magnitude at $x\sim
 \delta\sim\epsilon^2$ and only combinations
 $\delta^{2n}\epsilon^{2m}$ with $n\ge m$ are allowed, among
 which only $\delta^{2n}\epsilon^{2n}\sim W^{2n}$ remain finite
 at $\delta\to 0$.

 Landauer resistance $\rho$ is determined by a quadratic form of
 the matrix elements $\tau_{ij}$ \cite{6}. Consequently, growth
 exponents for $\langle \rho^n \rangle$  coincide with those for
 the $2n$th moments of $\tau_{ij}$.
 An expression for $\langle
 \rho^n \rangle$ contains a linear combination of the
 corresponding exponents, which leads to Eq. (2)
 if the complex-valued exponents are taken into account.

\end{document}